\documentclass[12pt]{article}%
\usepackage{amssymb}%
\usepackage{amsmath}%
\usepackage{amsfonts}%
\usepackage{graphicx}

\begin{document}

\title{Lattice statistical models for the nematic transitions in liquid-crystalline systems}
\author{E. S. Nascimento, A. P. Vieira, and S. R. Salinas\\Instituto de F\'{\i}sica\\Universidade de S\~{a}o Paulo\\S\~{a}o Paulo, SP, 05508-090}
\date{16 August, 2016}
\maketitle

\begin{abstract}
We investigate the connections between some simple Maier-Saupe lattice models,
with a discrete choice of orientations of the microscopic directors, and a
recent proposal of a two-tensor formalism to describe the phase diagrams of
nematic liquid-crystalline systems. This two-tensor proposal is used to
formulate the statistical problem in terms of fully-connected lattice
Hamiltonians, with the local nematic directors restricted to the Cartesian
axes. Depending on the choice of interaction parameters, we regain all of the
main features of the original mean-field two-tensor calculations. With a
standard choice of parameters, we obtain the well-known sequence of isotropic,
uniaxial, and biaxial nematic structures, with a Landau multicritical point.
With another suitably chosen set of parameters, we obtain two tricritical
points, according to some recent predictions of the two-tensor calculations.
The simple statistical lattice models are quite easy to work with, for all
values of parameters, and the present calculations can be carried out beyond
the mean-field level.

\end{abstract}

\section{Introduction}

In a number of recent investigations of the phase diagrams of nematic
liquid-crystalline systems, with a view to characterizing biaxial nematic
structures, we have introduced some simple lattice models with Maier-Saupe
pair interactions \cite{CLS2010, CVS2011, LSY2011, NHVS2015}. We claim that
these schematic statistical lattice models, with a restrict choice of
orientational variables, and adequately chosen interaction parameters, are
already able to account for the qualitative features of the complex phase
diagrams of nematic systems. Also, they are amenable to relatively easy
statistical mechanics calculations, for a wide choice of parameters, and these
calculations may be carried out beyond a mean-field scenario \cite{CVS2011}.
We were then motivated to investigate the connections of these statistical
formulations with some proposals of a quite general two-tensor formalism, at
the mean-field level, which describes the classical isotropic-uniaxial-biaxial
sequence of phases in the nematic systems as well as the onset of a
tricritical point for a particular choice of interaction parameters
\cite{SVD2003}\cite{MV2005}.

We first review the definitions of the \textquotedblleft order
parameters\textquotedblright\ and the general form of interaction energy used
by Sonnet, Virga, and Durand \cite{SVD2003}. This SVD interaction reproduces
an early proposal of Straley \cite{Straley1974} for the most general form of
the interaction between a pair of nematogenic units. In Section II, we use
this pair interaction to introduce a fully-connected lattice model, with a
restriction of the microscopic nematic directors to point along the Cartesian
axis, which is reminiscent of an old proposal of Zwanzig to treat the Onsager
model of rigid cylinders \cite{Zwanzig}. If we adopt this Zwanzig suppression
of microscopic fluctuations, the statistical problem is considerably
simplified. In Section III we describe some particular cases. For the simplest
choice of interaction parameters, which corresponds to the so-called MSZ
model, we regain the weak first-order transition between the isotropic and the
uniaxial nematic phases \cite{CVS2011}\cite{LS2014}. The inclusion of a
special form of extra (intrinsically biaxial) interactions, which has been
used by most investigators in this area, leads to the usual
isotropic-uniaxial-biaxial sequence, and to a Landau multicritical point
\cite{NHVS2015}. With this choice of parameters, we define an MSZ6 model,
which is equivalent to a lattice statistical model originally proposed by
Boccara, Medjani, and de S\`{e}ze \cite{BMS1977}. Although it is not our aim
in the present paper to offer a thorough review of the literature, present
results are in qualitative agreement with earlier numerical calculations for
lattice systems of non-cylindrically symmetric particles \cite{LR1980}%
\cite{LZNS1975}. With another particular choice of energy parameters, we do
confirm the existence of a pair of tricritical points, according to the
predictions of Matteis and Virga \cite{MV2005}.

\section{The two-tensor order parameter}

According to Sonnet, Virga, and Durand \cite{SVD2003}, the usual molecular
nematic tensor $\mathbf{q}$ is given by the $3\times3$ matrix,%
\begin{equation}
\mathbf{q}=\overrightarrow{n}_{1}\otimes\overrightarrow{n}_{1}-\frac{1}%
{3}\mathbf{I,}%
\end{equation}
where $\mathbf{I}$ is a unit matrix, the symbol $\otimes$ indicates a tensor
product, and $\overrightarrow{n}_{1}$, with $\left\vert \overrightarrow{n}%
_{1}\right\vert =1$, is the director associated with a uniaxial nematogenic
molecule. In a more explicit notation, we write%
\begin{equation}
\mathbf{q}=\left(
\begin{array}
[c]{ccc}%
n_{1x}n_{1x}-\frac{1}{3} & n_{1x}n_{1y} & n_{1x}n_{1z}\\
n_{1y}n_{1x} & n_{1y}n_{1y}-\frac{1}{3} & n_{1y}n_{1z}\\
n_{1z}n_{1x} & n_{1z}n_{1y} & n_{1z}n_{1z}-\frac{1}{3}%
\end{array}
\right)  ,
\end{equation}
which corresponds to the well-known traceless nematic tensor order parameter.
Also, we can write the compact form%
\begin{equation}
q^{\mu\nu}=\left(  n_{1\mu}n_{1\nu}-\frac{1}{3}\delta_{\mu\nu}\right)  ,
\end{equation}
where $\mu,\nu=x,y,z$.

In order to account for the biaxiality, Sonnet, Virga, and Durand consider two
additional unit vectors, $\overrightarrow{n}_{2}$ and $\overrightarrow{n}_{3}%
$, along the two minor axes of a nematogenic molecule, so that
$\overrightarrow{n}_{2}$ is normal to $\overrightarrow{n}_{3}$, and both of
them are normal to the director $\overrightarrow{n}_{1}$ along the major axis
($\overrightarrow{n}_{2}\cdot\overrightarrow{n}_{3}=0$ and $\overrightarrow
{n}_{1}\cdot\overrightarrow{n}_{2}=\overrightarrow{n}_{1}\cdot\overrightarrow
{n}_{3}=0$). We then define a second traceless tensor,%
\begin{equation}
\boldsymbol{b}=\overrightarrow{n}_{2}\otimes\overrightarrow{n}_{2}%
-\overrightarrow{n}_{3}\otimes\overrightarrow{n}_{3},
\end{equation}
which can be written as%
\begin{equation}
\mathbf{b}=\left(
\begin{array}
[c]{ccc}%
n_{2x}n_{2x}-n_{3x}n_{3x} & n_{2x}n_{2y}-n_{3x}n_{3y} & n_{2x}n_{2z}%
-n_{3x}n_{3z}\\
n_{2y}n_{2x}-n_{3y}n_{3x} & n_{2y}n_{2y}-n_{3y}n_{3y} & n_{2y}n_{2z}%
-n_{3y}n_{3z}\\
n_{2z}n_{2x}-n_{3z}n_{3x} & n_{2z}n_{2y}-n_{3z}n_{3y} & n_{2z}n_{2z}%
-n_{3z}n_{3z}%
\end{array}
\right)  .
\end{equation}
In a more compact form, we write%
\begin{equation}
b^{\mu\nu}=\left(  n_{2\mu}n_{2\nu}-n_{3\mu}n_{3\nu}\right)  ,
\end{equation}
from which it is easy to see that%
\begin{equation}
\operatorname*{Tr}\mathbf{b=}%
{\displaystyle\sum\limits_{\mu}}
b^{\mu\mu}=0,
\end{equation}
since $\overrightarrow{n}_{2}$ and $\overrightarrow{n}_{3}$ are unit vectors.

Consider a pair of nematogenic molecules, which are associated with the local
tensor parameters $\left(  \mathbf{q},\mathbf{b}\right)  $ and $\left(
\mathbf{q}^{\prime},\mathbf{b}^{\prime}\right)  $. We now write the
orientational interaction energy between these molecules,%
\begin{equation}
V=-U_{0}\left\{  \mathbf{q}\cdot\mathbf{q}^{\prime}+\gamma\left(
\mathbf{q}\cdot\mathbf{b}^{\prime}+\mathbf{b}\cdot\mathbf{q}^{\prime}\right)
+\lambda\,\mathbf{b}\cdot\mathbf{b}^{\prime}\right\}  ,\label{svdgeral}%
\end{equation}
where $U_{0}>0$ is a typical interaction energy parameter, and the
dimensionless parameters $\gamma$ and $\lambda$ may be arbitrarily chosen. If
$\gamma=\lambda=0$, we recover the interaction energy associated with the
usual form of the Maier-Saupe model, which is known to lead to a uniaxial
nematic phase. If either $\gamma\neq0$ or $\lambda\neq0$, the biaxial
components contribute to the interaction energy, and we may anticipate the
existence of a biaxial nematic phase. As it has been pointed out by Sonnet,
Virga, and Durand \cite{SVD2003}, this expression of $V$ is the most general
orientational form of the interaction energy between a pair of molecules if we
restrict to linear terms of the products of the tensor order parameters and
require invariance under the exchange of molecules. Also, this expression of
$V$ corresponds to the general form of the pair interaction energy that has
been proposed in the pioneering work of Straley \cite{Straley1974}.

According to the mean-field calculation of Sonnet, Virga, and Durand
\cite{SVD2003}, there is a wide range of acceptable values of the parameters
$\gamma$ and $\lambda$. For example, we can make the choice $\gamma=0$, which
eliminates the couplings between uniaxial and biaxial terms. The interaction
energy in this case reduces to the expression%
\begin{equation}
V=-U_{0}\left\{  \mathbf{q}\cdot\mathbf{q}^{\prime}+\lambda\,\mathbf{b}%
\cdot\mathbf{b}^{\prime}\right\}  .
\end{equation}
Besides the expected isotropic-uniaxial-biaxial sequence of phases, it has
been shown that there is a tricritical point in a phase diagram in terms of
$\lambda$ and the temperature $T$. In a more recent work, De Matteis and Virga
\cite{MV2005} have shown that there are indeed two distinct tricritical
points. In the present article, we show that calculations for the
corresponding lattice statistical model, with the suppression of most
orientational fluctuations, do lead to the same qualitative behavior.

An interesting choice of parameters, $\lambda=\gamma^{2}$, is related to the
geometric-mean approximation, which is suggested by an analysis of London%
\'{}%
s dispersion forces, and which is the most investigated case in the literature
(see, for example, the recent work by Luckhurst and collaborators
\cite{LNSTT2012}). The interaction energy in this case assumes the simple form%
\begin{equation}
V=-U_{0}\left(  \mathbf{q}+\gamma\mathbf{b}\right)  \cdot\left(
\mathbf{q}^{\prime}+\gamma\mathbf{b}^{\prime}\right)  .
\end{equation}
Again, the elementary lattice model reproduces all the qualitative findings of
more involved calculations.

\section{Statistical formulation of the SVD model}

Given the general form of the interaction energy between pairs of molecules,
eq. (\ref{svdgeral}), we write a fully-connected lattice Hamiltonian,%
\begin{equation}
\mathcal{H}=-\frac{U_{0}}{2N}\sum_{i,j=1}^{N}\sum_{\mu\nu}\left[  q_{i}%
^{\mu\nu}q_{j}^{\mu\nu}+\gamma\left(  q_{i}^{\mu\nu}b_{j}^{\mu\nu}+b_{i}%
^{\mu\nu}q_{j}^{\mu\nu}\right)  +\lambda b_{i}^{\mu\nu}b_{j}^{\mu\nu}\right]
,
\end{equation}
where $i,j=1,2,...N$, and $\mu,\nu=x,y,z$. It is then convenient to rewrite
this Hamiltonian in the form%
\begin{equation}
\mathcal{H}=-\frac{U_{0}}{2N}\sum_{\mu\nu}\left[  \left(  \sum_{i=1}^{N}%
q_{i}^{\mu\nu}\right)  ^{2}+2\gamma\left(  \sum_{i=1}^{N}q_{i}^{\mu\nu
}\right)  \left(  \sum_{i=1}^{N}b_{i}^{\mu\nu}\right)  +\lambda\left(
\sum_{i=1}^{N}b_{i}^{\mu\nu}\right)  ^{2}\right]  .
\end{equation}
In the Zwanzig-type models, with local directors restricted to the Cartesian
directions, as we explicitly indicate in Appendix A, this expression is
further simplified, since the local tensors do not have off-diagonal elements.
In this case, we have%
\begin{equation}
\mathcal{H}=-\frac{U_{0}}{2N}\sum_{\mu=1,2,3}\left[  \left(  \sum_{i=1}%
^{N}q_{i}^{\mu\mu}\right)  ^{2}+2\gamma\left(  \sum_{i=1}^{N}q_{i}^{\mu\mu
}\right)  \left(  \sum_{i=1}^{N}b_{i}^{\mu\mu}\right)  +\lambda\left(
\sum_{i=1}^{N}b_{i}^{\mu\mu}\right)  ^{2}\right]  .\label{hmsz6}%
\end{equation}

The canonical partition function of this problem is given by%
\begin{equation}
Z=\operatorname*{Tr}\exp\left\{  \frac{\beta U_{0}}{2N}\sum_{\mu}\left[
\left(  \sum_{i=1}^{N}q_{i}^{\mu\mu}\right)  ^{2}+2\gamma\left(  \sum
_{i=1}^{N}q_{i}^{\mu\mu}\right)  \left(  \sum_{i=1}^{N}b_{i}^{\mu\mu}\right)
+\lambda\left(  \sum_{i=1}^{N}b_{i}^{\mu\mu}\right)  ^{2}\right]  \right\}  ,
\end{equation}
where $\beta=1/T$ is the inverse of temperature $T$, and the trace indicates a
sum over the (microscopic) configurations of the tensors $\mathbf{q}$ and
$\mathbf{b}$. We now introduce a new parameter,%
\begin{equation}
\omega^{2}=\lambda-\gamma^{2},
\end{equation}
and write%
\begin{equation}
Z=\operatorname*{Tr}\exp\left\{  \sum_{\mu}\frac{\beta U_{0}}{2N}\left[
\sum_{i=1}^{N}\left(  q_{i}^{\mu\mu}+\gamma b_{i}^{\mu\mu}\right)  \right]
^{2}+\frac{\beta U_{0}}{2N}\omega^{2}\left[  \sum_{i=1}^{N}b_{i}^{\mu\mu
}\right]  ^{2}\right\}  .
\end{equation}
If $\omega^{2}>0$, it is quite standard to linearize the quadratic terms by
the introduction of two integral Gaussian identities. Otherwise, if
$\omega^{2}<0$, we can resort to integral representations of two Dirac delta
functions. Note the particular cases: (i) $\gamma=\lambda=0$, which
corresponds to the well-known uniaxial model; (ii) $\omega=0$, in other words,
$\lambda=\gamma^{2}$, which corresponds to the most widely used case (and to
the geometrical mean approximation of Luckhurst and collaborators
\cite{LSY2011}); (iii) $\gamma=0$, which has been analyzed in detail by Sonnet
and collaborators \cite{SVD2003}, and by De Matteis and Virga \cite{MV2005}.

We now resort to a set of Gaussian integral identities of the form%
\begin{equation}%
{\displaystyle\int\limits_{-\infty}^{+\infty}}
\frac{dx}{\sqrt{\pi}}\exp\left(  -x^{2}+2ax\right)  =\exp\left(  a^{2}\right)
.
\end{equation}
With a convenient choice of variables, and for $\omega^{2}>0$, it is easy to
write%
\[
Z=%
{\displaystyle\prod\limits_{\mu}}
\left(  \int\left(  \frac{\beta U_{0}N}{2\pi}\right)  ^{1/2}dx_{\mu}%
\int\left(  \frac{\beta U_{0}N}{2\pi}\right)  ^{1/2}dy_{\mu}\right)
\exp\left[  -\frac{1}{2}\beta U_{0}N%
{\displaystyle\sum\limits_{\mu}}
\left(  x_{\mu}^{2}+y_{\mu}^{2}\right)  \right]  \times
\]%
\begin{equation}
\times\left\{  \operatorname*{Tr}\exp\left[  \sum_{\mu}\beta U_{0}x_{\mu
}\left(  q^{\mu\mu}+\gamma b^{\mu\mu}\right)  +\sum_{\mu}\beta U_{0}x_{\mu
}\omega y_{\mu}b^{\mu\mu}\right]  \right\}  ^{N},
\end{equation}
where the single-site trace has to be calculated with the microscopic
configurations given in Appendix A. If we carry out the sum over these
accessible microscopic configurations, it is straightforward to show that%
\begin{equation}
Z=\int\left[  dx_{\mu}dy_{\mu}\right]  \exp\left[  -\frac{1}{2}\beta U_{0}N%
{\displaystyle\sum\limits_{\mu}}
\left(  x_{\mu}^{2}+y_{\mu}^{2}\right)  +N\ln2-\frac{1}{2}\beta U_{0}N%
{\displaystyle\sum\limits_{\mu}}
x_{\mu}+N\ln\mathcal{Z}_{0}\right]  ,
\end{equation}
with%
\[
\mathcal{Z}_{0}=\exp\left(  \beta U_{0}x_{1}\right)  \cosh\left[  \beta
U_{0}\gamma\left(  x_{2}-x_{3}\right)  +\beta U_{0}\omega\left(  y_{2}%
-y_{3}\right)  \right]  +
\]%
\[
+\exp\left(  \beta U_{0}x_{2}\right)  \cosh\left[  \beta U_{0}\gamma\left(
x_{1}-x_{3}\right)  +\beta U_{0}\omega\left(  y_{1}-y_{3}\right)  \right]  +
\]%
\begin{equation}
+\exp\left(  \beta U_{0}x_{3}\right)  \cosh\left[  \beta U_{0}\gamma\left(
x_{1}-x_{2}\right)  +\beta U_{0}\omega\left(  y_{1}-y_{2}\right)  \right]  .
\end{equation}
Therefore, we write%
\begin{equation}
Z=\int\left[  dx_{\mu}dy_{\mu}\right]  \exp\left[  -\beta U_{0}N\,f\right]  ,
\end{equation}
and minimize the density of free energy $f$ with respect to the variables
$\left\{  x_{\mu}\right\}  $ and $\left\{  y_{\mu}\right\}  $.

From the saddle-point equations,%
\begin{equation}
\frac{\partial f}{\partial x_{\mu}}=0;\qquad\frac{\partial f}{\partial y_{\mu
}}=0,
\end{equation}
we show that%
\begin{equation}
\sum_{\mu}x_{\mu}=\sum_{\mu}y_{\mu}=0,
\end{equation}
which is the traceless property of the tensors $\boldsymbol{x}$ and
$\boldsymbol{y}$. We thus have to analyze a system of just four equations and
four unknowns, and it is not difficult to guess some acceptable solutions.

From the physical point of view, this problem is more appealing if we
introduce the change of variables,%
\begin{equation}
x_{\mu}=Q_{\mu}+\gamma B_{\mu}%
\end{equation}
and%
\[
y_{\mu}=\omega B_{\mu}.
\]
In terms of these new variables, we have%
\begin{equation}
Z=\int\left[  dQ_{\mu}dB_{\mu}\right]  \exp\left[  -\beta U_{0}N\,f\right]  ,
\end{equation}
so that%
\begin{equation}
f=\frac{1}{2}\sum_{\mu}\left[  Q_{\mu}^{2}+2\gamma Q_{\mu}B_{\mu}+\lambda
B_{\mu}^{2}\right]  -\frac{1}{\beta U_{0}}\ln2-\frac{1}{3}\sum_{\mu}\left(
Q_{\mu}+\gamma B_{\mu}\right)  -\frac{1}{\beta U_{0}}\ln\widetilde
{\mathcal{Z}_{0}},
\end{equation}
with%
\[
\widetilde{\mathcal{Z}_{0}}=\exp\left[  \beta U_{0}\left(  Q_{1}+\gamma
B_{1}\right)  \right]  \cosh\left\{  \beta U_{0}\left[  \gamma\left(
Q_{2}-Q_{3}\right)  +\lambda\left(  B_{2}-B_{3}\right)  \right]  \right\}  +
\]%
\[
+\exp\left[  \beta U_{0}\left(  Q_{2}+\gamma B_{2}\right)  \right]
\cosh\left\{  \beta U_{0}\left[  \gamma\left(  Q_{1}-Q_{3}\right)
+\lambda\left(  B_{1}-B_{3}\right)  \right]  \right\}  +
\]%
\begin{equation}
+\exp\left[  \beta U_{0}\left(  Q_{3}+\gamma B_{3}\right)  \right]
\cosh\left\{  \beta U_{0}\left[  \gamma\left(  Q_{1}-Q_{2}\right)
+\lambda\left(  B_{1}-B_{2}\right)  \right]  \right\}  .
\end{equation}
Now it is relatively easy to obtain the minima of the function $f$ and to
check the traceless property,%
\begin{equation}
\sum_{\mu}Q_{\mu}=\sum_{\mu}B_{\mu}=0.
\end{equation}
These expressions are particularly convenient to be used for writing an
expansion of the free energy in the vicinity the isotropic phase and comparing
with a general Landau-de Gennes phenomenological expansion.

\section{Analyses of some special cases}

We now consider some special cases of our general expressions. As we have
mentioned in the Introduction, we first show that the simplest choice of
interaction parameters, corresponding to the simple MSZ\ model, leads to the
weak first-order transition between an isotropic and a uniaxial nematic phase.
We then analyze a special choice of parameters that leads to a phase diagram
with a rich tricritical behavior. Finally, we consider a choice of parameters
according to the usual geometric mean approximation, which leads to the
well-known isotropic-uniaxial-biaxial sequence of phases, and to a Landau
multicritical point.

\subsection{Uniaxial nematic transition}

The uniaxial nematic transition may be described by the simple pair
interaction energy%
\begin{equation}
V_{u}=-U_{0}\,\mathbf{q}\cdot\mathbf{q}^{\prime}=-U_{0}\sum_{\mu\nu}q^{\mu\nu
}\left(  q^{\prime}\right)  ^{\mu\nu},
\end{equation}
which can also be written as%
\begin{equation}
V_{u}=-U_{0}\,\left(  \overrightarrow{n}_{1}.\overrightarrow{n}_{1}^{\prime
}\right)  ^{2}+U_{0}.
\end{equation}
This expression corresponds to the uniaxial term of the general form of
interaction energy proposed by Straley (see the general form of interaction in
Appendix B).

This special case corresponds to the standard Maier-Saupe model, which has
been analyzed by numerous investigators. It leads to the usual weak
first-order transition between the isotropic and the uniaxial nematic
structure. In previous articles, we have shown that the restricted choice of
the microscopic degrees of freedom, as described in Appendix A, leads to the
definition of the Maier-Saupe-Zwanzig, or MSZ, lattice statistical model, with
the typical thermodynamic behavior of any simple interactions of uniaxial type.

\subsection{Particular case with $\gamma=0$\textbf{\ and }$\lambda\neq0$}

This particular case has been analyzed in detail by Sonnet, Virga, and Durand
\cite{SVD2003}, who pointed out the existence of a novel tricritical behavior.
Note that we can write%
\[
\mathbf{b}\cdot\mathbf{b}^{\prime}=\sum_{\mu\nu}\left(  n_{2\mu}n_{2\nu
}-n_{3\mu}n_{3\nu}\right)  \left(  n_{2\mu}^{\prime}n_{2\nu}^{\prime}-n_{3\mu
}^{\prime}n_{3\nu}^{\prime}\right)  =
\]%
\begin{equation}
=\left(  \overrightarrow{n}_{2}.\overrightarrow{n^{\prime}}_{2}\right)
^{2}-\left(  \overrightarrow{n}_{2}.\overrightarrow{n^{\prime}}_{3}\right)
^{2}-\left(  \overrightarrow{n}_{3}.\overrightarrow{n^{\prime}}_{2}\right)
^{2}+\left(  \overrightarrow{n}_{3}.\overrightarrow{n^{\prime}}_{3}\right)
^{2},
\end{equation}
which corresponds to the last term in the biaxial contribution to Straley%
\'{}%
s interaction energy (see Appendix B). Therefore, if $\gamma=0$\textbf{\ }and
$\lambda\neq0$, we have%
\[
V_{\lambda}=-U_{0}\,{\LARGE \{}\left(  \overrightarrow{n}_{1}.\overrightarrow
{n^{\prime}}_{1}\right)  ^{2}-1+
\]%
\begin{equation}
+\lambda\left[  \left(  \overrightarrow{n}_{2}.\overrightarrow{n^{\prime}}%
_{2}\right)  ^{2}-\left(  \overrightarrow{n}_{2}.\overrightarrow{n^{\prime}%
}_{3}\right)  ^{2}-\left(  \overrightarrow{n}_{3}.\overrightarrow
{n_{2}^{\prime}}\right)  ^{2}+\left(  \overrightarrow{n}_{3}.\overrightarrow
{n^{\prime}}_{3}\right)  ^{2}\right]  {\LARGE \}.}%
\end{equation}

In this case, it is convenient to write the lattice Hamiltonian%
\[
\mathcal{H}_{\lambda}=-U_{0}\frac{1}{2N}\sum_{i,j=1}^{N}\sum_{\mu\nu}\left[
q_{i}^{\mu\nu}q_{j}^{\mu\nu}+\lambda b_{i}^{\mu\nu}b_{j}^{\mu\nu}\right]  =
\]%
\begin{equation}
=-U_{0}\frac{1}{2N}\sum_{\mu\nu}\left[  \left(  \sum_{i}q_{i}^{\mu\nu}\right)
^{2}+\lambda\left(  \sum_{i}b_{i}^{\mu\nu}\right)  ^{2}\right]  ,
\end{equation}
which is associated with a partition function that may be easily simplified by
the use of two Gaussian integral identities. Calculations are straightforward,
and lead to a free energy functional and a set of equations of state that can
be numerically analyzed in great detail.

In figure 1 we draw a phase diagram in terms of the parameter $\lambda$ and
the temperature $T$ (in units of energy $U_{0}$). We indicate isotropic ($I$),
uniaxial ($U$) and biaxial ($B$) nematic regions. Broken lines indicate
first-order transitions. The solid line represents the second-order transition
between uniaxially and biaxially ordered nematic phases. In the inset of this
figure we show a tricritical point (at the end of the second-order line) and a
triple point (at the meeting of three first-order boundaries). This phase
diagram qualitatively agrees with the predictions of the calculations by
Sonnet and collaborators \cite{SVD2003}.

We now refer to figure 2, which is an amplification of figure 1 for
$0.4\lesssim\lambda\lesssim0.8$. We clearly see a second tricritical point, in
agreement with calculations of De Matteis and Virga \cite{MV2005}. Therefore,
although it may be difficult to experimentally characterize the tricritical
behavior, we show that the simple statistical lattice model reproduces the
main findings of the recent (and much more involved) calculations for the
phase diagram with this particular choice of energy parameters.%

\begin{figure}
[ptb]
\begin{center}
\includegraphics[
height=2.5391in,
width=3.9081in
]%
{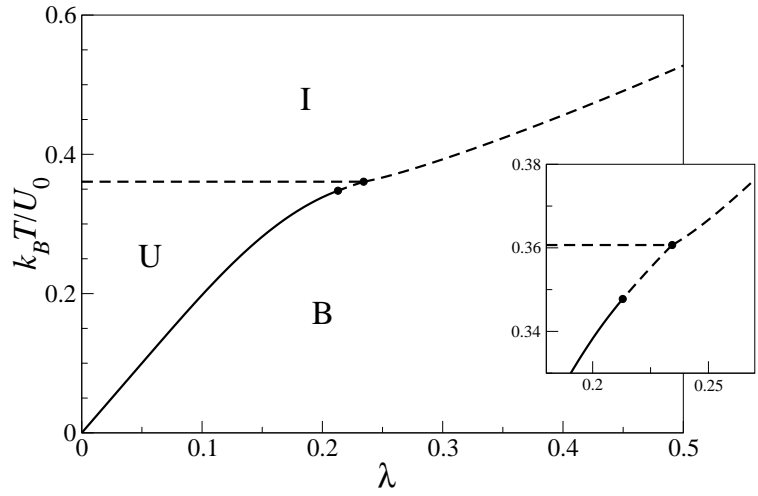}%
\caption{Phase diagram in terms of temperature $T$ (in convenient units) and
parameter $\lambda$, with $\gamma=0$. Dashed lines indicate first-order
transitions (the solid line is a second-order transition). We also indicate
nematic uniaxial (U), nematic biaxial (B) and isotropic (I) regions. This
diagram corresponds to the particular case analyzed by De Matteis and Virga.}%
\end{center}
\end{figure}
%

\begin{figure}
[ptb]
\begin{center}
\includegraphics[
height=2.5391in,
width=3.6659in
]%
{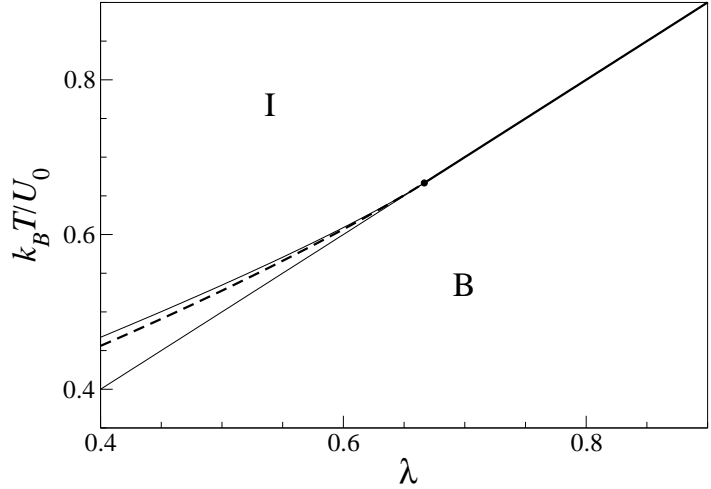}%
\caption{Second tricritical point (for $\lambda>0.6$) in the $\lambda-T$ phase
diagram with $\gamma=0$. The first-order boundary (dashed line) meets
tangentially with the critical boundary (solid line) at a tricritical point.
We also draw the stability lines of the isotropic and uniaxial nematic
solutions.}%
\end{center}
\end{figure}

\subsection{Geometric mean approximation}

Most of the theoretical investigations of biaxial nematic structures use a
pair interaction of the form%
\[
V_{l}=-U_{0}\left[  \mathbf{q}+\gamma\mathbf{b}\right]  .\left[
\mathbf{q}^{\prime}+\gamma\mathbf{b}^{\prime}\right]  =
\]%
\begin{equation}
=-U_{0}{\LARGE \{}\mathbf{q.q}^{\prime}+\gamma\left(  \mathbf{q.b}^{\prime
}+\mathbf{b.q}^{\prime}\right)  +\gamma^{2}\mathbf{b.b}^{\prime}{\LARGE \}},
\end{equation}
where $\gamma$ is a parameter of biaxiality. From the general expression of
eq. (\ref{svdgeral}), we have%
\begin{equation}
\lambda=\gamma^{2}\quad\Longrightarrow\quad\omega^{2}=\lambda-\gamma^{2}=0,
\end{equation}
which corresponds to the geometric mean approximation of Luckhurst and
collaborators (and to the simple MSZ6 model of our own previous article
\cite{NHVS2015}).

According to the calculations of Section 2, it is convenient to define%
\begin{equation}
\mathbf{Q}=\mathbf{q}+\gamma\,\mathbf{b,}%
\end{equation}
and write the usual quadratic energy for a Hamiltonian with pair interactions.
In the language of the six-state model (as in Appendix A), it is easy to see
that%
\[
\mathbf{Q}_{1}=\left(
\begin{array}
[c]{ccc}%
2/3 & 0 & 0\\
0 & -1/3 & 0\\
0 & 0 & -1/3
\end{array}
\right)  +\gamma\left(
\begin{array}
[c]{ccc}%
0 & 0 & 0\\
0 & 1 & 0\\
0 & 0 & -1
\end{array}
\right)  =
\]%
\begin{equation}
=\left(
\begin{array}
[c]{ccc}%
2/3 & 0 & 0\\
0 & -1/3+\gamma & 0\\
& 0 & -1/3-\gamma
\end{array}
\right)  ,
\end{equation}
with the additional five tensors, $\mathbf{Q}_{2}$ to $\mathbf{Q}_{6}$, given
by permutations of the diagonal elements.

In figure 3, we show a phase diagram that has been obtained by many authors
(see, for example, the recent review of computer simulations by Berardi and
coworkers \cite{BMORZ2008}).%

\begin{figure}
[ptb]
\begin{center}
\includegraphics[
height=2.5391in,
width=3.6305in
]%
{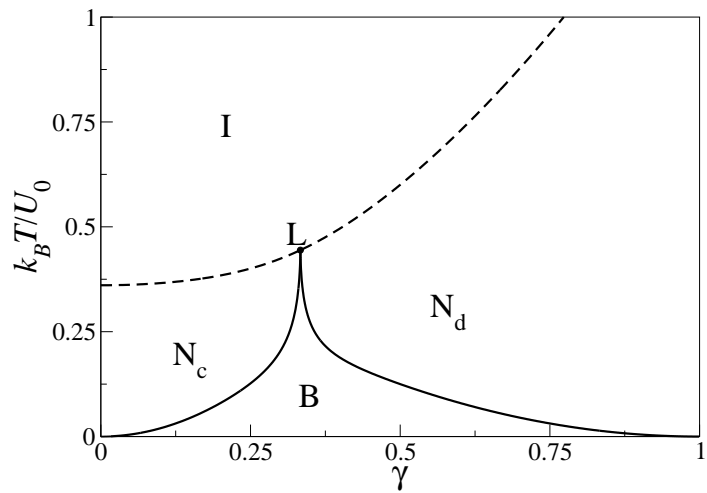}%
\caption{Phase diagram in terms of temperature $T$ (in convenient units) and
the parameter $\gamma$, with $\lambda=\gamma^{2}$, which corresponds to the
usual geometric mean approximation. We indicate the isotropic ($I$), biaxial
($B$), and two distinct uniaxial nematic phases ($N_{c}$ and $N_{d}$). The
dashed line is a first-order transition. Solid lines are second-order
transitions that meet at the Landau multicritcial point ($L$).}%
\end{center}
\end{figure}

\section{Conclusions}

We have used a general form of the interaction energy between pairs of
molecules to introduce a class of fully-connected elementary lattice models.
If the microscopic nematic directors are restricted to point along the
Cartesian axis, the statistical problem is amenable to standard calculations.
We have used this approach to account for the main features of complex phase
diagrams, in terms of temperature and a wide choice of model parameters. With
the simplest choice of parameters, we reproduce the well-known weak
first-order transition between the isotropic and the uniaxial nematic phases.
With a choice of interaction parameters according to a geometric mean rule,
which has been used by a large number of investigators, we reproduce the
well-known sequence of isotropic to uniaxial nematic and then to a biaxial
nematic phase as a function of decreasing temperatures. Also, we locate a
Landau multicritical point, and can easily make contact with the standard
Landau-de Gennes phenomenological expansions of the free energy, with the
advantage of calculating the expansion coefficients in terms of the model
parameters. For a special choice of parameters, we reproduce a phase diagram
with a rich tricritical behavior, as it has been pointed out in the recent
literature. We claim that these elementary lattice models can be easily used
to explore new situations and to go beyond the simplest mean-field calculations.

\begin{center}
\textbf{Appendix A\bigskip}
\end{center}

In the six-state model, we have the following microscopic configurations of
the local nematic directors:%

\[
(i)\quad\overrightarrow{n}_{1}=\left(  \pm1,0,0\right)  ;\quad\overrightarrow
{n}_{2}=\left(  0,\pm1,0\right)  ;\quad\overrightarrow{n}_{3}=\left(
0,0,\pm1\right)
\]%
\[
(ii)\quad\overrightarrow{n}_{1}=\left(  \pm1,0,0\right)  ;\quad\overrightarrow
{n}_{2}=\left(  0,0,\pm1\right)  ;\quad\overrightarrow{n}_{3}=\left(
0,\pm1,0\right)
\]%
\[
(iii)\quad\overrightarrow{n}_{1}=\left(  0,\pm1,0\right)  ;\quad
\overrightarrow{n}_{2}=\left(  \pm1,0,0\right)  ;\quad\overrightarrow{n}%
_{3}=\left(  0,0,\pm1\right)
\]%
\[
(iv)\quad\overrightarrow{n}_{1}=\left(  0,\pm1,0\right)  ;\quad\overrightarrow
{n}_{2}=\left(  0,0,\pm1\right)  ;\quad\overrightarrow{n}_{3}=\left(
\pm1,0,0\right)
\]%
\[
(v)\quad\overrightarrow{n}_{1}=\left(  0,0,\pm1\right)  ;\quad\overrightarrow
{n}_{2}=\left(  \pm1,0,0\right)  ;\quad\overrightarrow{n}_{3}=\left(
0,\pm1,0\right)
\]%
\[
(vi)\quad\overrightarrow{n}_{1}=\left(  0,0,\pm1\right)  ;\quad\overrightarrow
{n}_{2}=\left(  0,\pm1,0\right)  ;\quad\overrightarrow{n}_{3}=\left(
\pm1,0,0\right)
\]

\bigskip

\begin{center}
\textbf{Appendix B}\bigskip
\end{center}

The most general form of the orientational interaction energy between two
biaxial objects is given by the Straley formula,%
\[
V=\alpha+\frac{1}{2}\beta\left[  \left(  \overrightarrow{n}_{1}%
.\overrightarrow{n^{\prime}}_{1}\right)  ^{2}-1\right]  +
\]%
\[
+2\gamma\left[  \left(  \overrightarrow{n}_{3}.\overrightarrow{n^{\prime}}%
_{3}\right)  ^{2}-\left(  \overrightarrow{n}_{2}.\overrightarrow{n^{\prime}%
}_{2}\right)  ^{2}\right]  +
\]%
\begin{equation}
+\frac{1}{2}\delta\left[  \left(  \overrightarrow{n}_{2}.\overrightarrow
{n^{\prime}}_{2}\right)  ^{2}-\left(  \overrightarrow{n}_{2}.\overrightarrow
{n^{\prime}}_{3}\right)  ^{2}-\left(  \overrightarrow{n}_{3}.\overrightarrow
{n_{2}^{\prime}}\right)  ^{2}+\left(  \overrightarrow{n}_{3}.\overrightarrow
{n^{\prime}}_{3}\right)  ^{2}\right]  .\label{B1}%
\end{equation}
According to Straley \cite{Straley1974}, this expression is obtained under the
following assumptions: (i) it is restricted to quadratic terms of sines and
cosines of the relative directions; (ii) it is invariant under the inversion
of the axes; (iii) it is symmetric under the exchange of particles.

Also, according to the work of Straley, we can write%
\begin{equation}
V=\alpha+\beta F_{1}\left(  \theta\right)  +\gamma\left[  F_{2}\left(
\theta,\phi\right)  +F_{3}\left(  \theta,\psi\right)  \right]  +\delta
F_{4}\left(  \theta,\phi,\psi\right)  ,
\end{equation}
in terms of four basis functions and the Euler angles $\left(  \theta
,\phi,\psi\right)  $. If we restrict the calculations to molecules with the
symmetry of a rectangular parallelepiped, the associated distribution
functions are invariant under the transformations%
\begin{align}
\phi & \rightarrow\phi+\pi;\\
\psi & \rightarrow\psi+\pi;\nonumber\\
\theta & \rightarrow\pi-\theta;\quad\phi\rightarrow\pi-\phi;\quad
\psi\rightarrow-\psi,\nonumber
\end{align}
so that we have%
\begin{align}
F_{1}  & =\frac{1}{2}\left(  3\cos^{2}\theta-1\right)  ;\qquad F_{2}%
=\operatorname{sen}^{2}\theta\cos2\phi;\qquad F_{3}=\operatorname{sen}%
^{2}\theta\cos2\psi;\\
F_{4}  & =\frac{1}{2}\left(  1+\cos^{2}\theta\right)  \cos2\phi\cos2\psi
-\cos\theta\operatorname{sen}2\phi\operatorname{sen}2\psi.\nonumber
\end{align}

The general form of interaction of Sonnet, Virga, and Durand, which can be
written as%
\[
V=-U_{0}\,\,{\LARGE \{}{\Huge \,}\left(  \overrightarrow{n}_{1}%
.\overrightarrow{n^{\prime}}_{1}\right)  ^{2}-1+
\]%
\[
+\gamma\left[  \left(  \overrightarrow{n}_{1}.\overrightarrow{n^{\prime}}%
_{2}\right)  ^{2}-\left(  \overrightarrow{n}_{1}.\overrightarrow{n^{\prime}%
}_{3}\right)  ^{2}+\left(  \overrightarrow{n}_{2}.\overrightarrow{n^{\prime}%
}_{1}\right)  ^{2}-\left(  \overrightarrow{n}_{3}.\overrightarrow{n^{\prime}%
}_{1}\right)  ^{2}\right]  +
\]%
\begin{equation}
+\lambda\left[  \left(  \overrightarrow{n}_{2}.\overrightarrow{n^{\prime}}%
_{2}\right)  ^{2}-\left(  \overrightarrow{n}_{2}.\overrightarrow{n^{\prime}%
}_{3}\right)  ^{2}-\left(  \overrightarrow{n}_{3}.\overrightarrow
{n_{2}^{\prime}}\right)  ^{2}+\left(  \overrightarrow{n}_{3}.\overrightarrow
{n^{\prime}}_{3}\right)  ^{2}\right]  \,{\Huge \}\,}{\LARGE ,}%
\end{equation}
has been shown to be equivalent to the Straley expression \cite{SVD2003}.

\end{document}